\documentclass[english, preprint,12pt]{elsarticle}
\usepackage{amssymb}
\usepackage{array}
\usepackage{units}
\usepackage{graphicx}

\usepackage{mathrsfs}

\usepackage{epsfig}
\usepackage{natbib}
\usepackage{bm}

\usepackage{morefloats}

\makeatother

\usepackage{babel}

\makeatother

\usepackage{babel}
\usepackage{color}

\begin{document}

\begin{frontmatter}

\title{Quantum criticality of the Ohmic spin-boson model in a high dense spectrum: symmetries,quantum fluctuations and correlations}

\author{Xiaohui Qian}
\author{Congzhi Zeng}
\author{Nengji Zhou\corref{cor}}
\cortext[cor]{Corresponding author.}
\ead{zhounengji@hznu.edu.cn}

\address{Department of Physics, Hangzhou Normal University, Hangzhou 311121, China, PRC}


\begin{abstract}
Study of dissipative quantum phase transitions in the Ohmic spin-boson model is numerically challenging in a dense limit of environmental modes. In this work, large-scale numerical simulations are carried out based on the variational principle. The validity of variational calculations, spontaneous breakdown of symmetries, and quantum fluctuations and correlations in the Ohmic bath are carefully analyzed, and the critical coupling as well as exponents are accurately determined in the weak tunneling and continuum limits. In addition, quantum criticality of the Ohmic bath is uncovered both in the delocalized phase and at the transition point.
\end{abstract}

\end{frontmatter}

\section{Introduction}
Quantum phase transitions have been under intensive study over many decades in various correlated matters and light-matter interacting systems \cite{wei07,hur10,sac11}. The accurate description of quantum effects is essential to the understanding of such quantum critical phenomena. As a paradigmatic minimal example, the spin-boson model (SBM) consisting of a spin-$1/2$ particle (two-level system) and a bosonic environment has attracted significant interest \cite{leg87,hur08,bre16}. In spite of apparent simplicity, it catches the physics of a large range of different physical systems going from defects in solids and quantum thermodynamics \cite{lew88,gol92} to physical chemistry and biological systems \cite{cha95,eng07,col09}. It has also been used to study spontaneous emission in quantum optics \cite{gar17}, semiconducting quantum dots in nanocavities \cite{ota11}, trapped ions \cite{por08}, quantum heat engines \cite{uzd15}, and superconducting circuits \cite{lep18}. The ground-state and dynamic properties of SBM have been extensively and persistently investigated with analytical and numerical approaches \cite{sil84,chi11,naz12,ber14,wu17,pin18}. In particular, the localized-delocalized ground-state transition and coherent-incoherent dynamic transition have been detected with the increase of the system-environment coupling \cite{hur08,leg87,nal13}. Besides, many activities have also been devoted to the variants of SBM for richer phase diagrams \cite{guo12,zhou14,zho18,wan20}.

As the most well-known case, the Ohmic SBM has a linear spectral density function $J(\omega)\sim \alpha \omega^s$ with $s=1$ to characterize the coupling between the system and the environment. The model can be mapped onto the anisotropic Kondo model and interacting resonant level model based on the equivalence between Fermi and Bose operators in one dimension \cite{gui85}. Thus, the localized-delocalized phase transition of the Kosterlitz-Thouless type has been predicted, and the critical coupling is located around $\alpha = 1$ associated with the discontinuous jumps of the spin magnetization and entropy \cite{hur08}. Different from the single-spin case, however, there was much debate among numerical works concerning the value of the critical coupling $\alpha_c$ for the two-impurity model, due to the lack of the analytical solution \cite{ort10,mcc10,win14,zho18}. Therefore, accurate determination of the transition point for the Ohmic SBM is still needed in numerical work to provide the methodological benchmark. Besides, the Ohmic SBM has been realized in recent experiments of superconducting quantum circuits wherein the spectral width of the reservoir is restricted \cite{mag18,lep18}. But the influence of the frequency range on the critical coupling $\alpha_c$ is still an open question.

A variety of numerical approaches have been employed to determine the nature of localized-delocalized transition and exact value of the critical coupling, e.g., numerical renormalization group (NRG), exact diagonalization, variational matrix product states, density-matrix renormalization group, quantum Monte Carlo (QMC), and variational method \cite{voj05,alv09,win09,zha10,guo12,ber14}. While numerical results of critical couplings show considerable differences in the shallow sub-Ohmic regime with $s > 0.5$. For instance, the NRG value of $\alpha_c$ is greater than others by nearly $10$ percent at $s=0.9$, let alone the Ohmic case $s=1$ \cite{win09,zha10,alv09}. Possible reason for the deviation is the numerical sensitivity of the phase transition nearby $s=1$.

Furthermore, numerical calculations are exact only in the continuum limit corresponding to a high dense spectrum. In that case, however, the scale separation breaks down, and the truncation becomes unmanageable \cite{ber14}.  Accordingly, the linear extrapolation was used to determine the value of transition point \cite{bul05}. But the linear dependence on the discretization parameter seems less convincing in the high dense spectrum. Very recently, quantum phase transitions of the Ohmic SBM in the continuum limit have been explored with the imaginary-time propagation \cite{wan19, fil20}. In spite of the critical coupling $\alpha \rightarrow 1^{+}$ has been arrived at directly, a detailed understanding of symmetries and quantum criticality of the Ohmic bath has been still lacking.

The pioneer variational work of the Ohmic SBM was based on the polaronic unitary transformation proposed by Silbey and Harris \cite{sil84}. Later on, the variational polaron ansatz was improved by superposing more than one coherent states and removing the imposed symmetry constrain \cite{chi11,naz12,zhe15,flo15,he18}. Recently, numerical variational method (NVM) has been developed based on systematic coherent-state decomposition of many-body ground state \cite{zhou14,blu17}. Excellent accuracy and reliability of the NVM have been proved in tackling ground-state phase transitions and quantum dynamics in the sub-Ohmic regime \cite{zhou15,zhou15b,zhou16,wan16, wan17}. However, the validity of the variational method for the Ohmic phase transition has not yet been demonstrated in the case of a high dense spectrum. Moreover, the attention in previous studies was mainly focused on the spin-related observations, especially for the spontaneous spin magnetization. In fact, bath observables provide a direct measurement of the quantum criticality intrinsic to the environment possessing many-body effects. But critical behaviors of quantum fluctuations and correlations in the Ohmic bath have not been clearly addressed so far.

In this article, quantum fluctuations and correlations in the Ohmic bath as well as the mechanism of spontaneous symmetry breaking are investigated with NVM for the Kosterlitz-Thouless transition. The transition point and exponents are accurately determined, and the validity of variational calculations is carefully examined. The rest of the paper is organized as follows. In section~\ref{sec:mod}, the model and variational approach are described. In section~\ref{sec:num}, numerical results are presented for the spontaneous breakdown of symmetries, the characteristic of the ground-state wavefunction, and the quantum criticality of the Ohmic bath. Finally, conclusions are drawn in section~\ref{sec:con}.

\section{Model and Method}\label{sec:mod}
The standard Hamiltonian of SBM can be written as
\begin{equation}
\label{Ohami}
\hat{H} =  \frac{\varepsilon}{2}\sigma_z-\frac{\Delta}{2}\sigma_x + \sum_{k} \omega_k b_{k}^\dag b_{k}  +  \frac{\sigma_z}{2}\sum_k \lambda_k(b^\dag_{k}+b_{k}),
\end{equation}
where $\varepsilon$ ($\Delta$) denotes the energy bias (bare tunneling amplitude), $b^\dag_k$ ($b_k$) is the bosonic creation (annihilation) operator of the $k$-th bath mode with the frequency $\omega_k$, $\sigma_x$ and $\sigma_z$ represent the Pauli spin-$1/2$ operators, and $\lambda_k$ signifies the coupling amplitude between the system and environment. With the coarse-grained treatment based on the Wilson energy mesh \cite{bul05, voj05,zha10,zhou14, blu17}, the values of $\lambda_k$ and  $\omega_k$ can be calculated by the continuous spectral density function $J(\omega)=2\alpha\omega_c^{1-s}\omega^{s}\Theta(\omega_c-\omega) =\sum_k\lambda_k^2\delta(\omega-\omega_k)$ after partitioning the phonon frequency domain $[0, \omega_c]$ into $M$ intervals $[\Lambda_k, \Lambda_{k+1}]\omega_c$ ($k=0, 1, \ldots, M-1$),
\begin{equation}
\label{sbm1_dis}
\lambda_k^2  =  \int^{\Lambda_{k+1}\omega_c}_{\Lambda_k\omega_c}dt J(t), \quad \omega_k  =  \lambda^{-2}_k \int^{\Lambda_{k+1}\omega_c}_{\Lambda_k\omega_c}dtJ(t)t,
\end{equation}
where $M$ is the number of effective bath modes, and $\Theta(\omega_c-\omega)$ is the Heaviside step function. To simplify notations, hereafter we fix the Planck constant $\hbar=1$ and the maximum frequency in the bath $\omega_c=1$. Other model parameters, i.e., $\varepsilon, \Delta$, and $\alpha$, are then set to be dimensionless. A logarithmic discretization procedure with the parameter $\Lambda_k=\Lambda^{k-M}$ is usually adopted \cite{bul03,hur08,chi11,fre13},  and the Wilson parameter $\Lambda \rightarrow 1$ is required for the Ohimc SBM (i.e., $s=1$) in order to obtain an accurate quantum criticality of the Kosterlitz-Thouless transition. However, $\Lambda=1.4\sim2.0$ was used in earlier numerical works \cite{bul05,ort10,ber14} where the critical coupling deviates from the prediction $\alpha_c=1$ by more than $10$ percent due to the finite size effect. In this paper, main results are presented with $\Lambda=1.01$. Additional simulations with $\Lambda=1.02$ confirm that the effect of discretization is already sufficiently small.

In variational calculations, a systematic coherent-state expansion is used \cite{zhou14,zhou15,wan16,blu17,zho18},
\begin{eqnarray}
\label{vmwave}
|\Psi \rangle & = & | \uparrow \rangle \sum_{n=1}^{N} A_n \exp\left[ \sum_{k=1}^{M}\left(f_{n,k}b_k^{\dag} - \mbox{H}.\mbox{c}.\right)\right] |0\rangle_{\textrm{b}} \nonumber \\
              & + & |\downarrow \rangle \sum_{n=1}^{N} D_n \exp\left[ \sum_{k=1}^{M}\left(g_{n,k}b_k^{\dag} - \mbox{H}.\mbox{c}.\right)\right] |0\rangle_{\textrm{b}},
\end{eqnarray}
where H$.$c$.$ denotes Hermitian conjugate, $\uparrow$ ($\downarrow$) stands for the spin up (down) state, and $|0\rangle_{\rm b}$ is the vacuum state of the bosonic bath. The variational parameters $f_{n,k}$ and $g_{n,k}$ represent the displacements of the coherent states  correlated
to the spin configurations $|\uparrow\rangle$ and $|\downarrow\rangle$, respectively,  and $A_n$ and $D_n$ are weights of the coherent states. The subscripts $n$ and $k$ correspond to the ranks of the coherent superposition state and effective bath mode, respectively. The energy can be then expressed as $E=\mathcal{H}/\mathcal{N}$ using the Hamiltonian expectation $\mathcal{H}=\langle \Psi|\hat{H}|\Psi\rangle$ and norm of the wave function $\mathcal{N}=\langle \Psi |\Psi\rangle$. By minimizing the energy to search for the ground state $|\Psi_g\rangle$, the variational procedure entails a set of self-consistency equations
\begin{equation}
\label{vmit}
\frac{\partial \mathcal{H}}{\partial x_{i}} - E\frac{\partial \mathcal{N}}{\partial x_{i}} = 0,
\end{equation}
where $x_i$ is a certain variational parameter $f_{n,k},g_{n,k},A_n$, or $D_n$. For each set of the model parameters $(\alpha, M, \Lambda, \varepsilon)$, more than one hundred random initial states are used in simulations to find the ground state. Furthermore, simulated annealing algorithm is also employed to escape from metastable states.

Besides the ground-state energy $E_g$ as well as the spin magnetization $\langle \sigma_{z} \rangle=\langle \Psi_{\rm g}|\sigma_{z}|\Psi_{\rm g} \rangle$, other observables related to the Ohmic bath are also investigated in the study of quantum phase transitions, which are the variances of phase-space variables $\Delta X_{\rm b}$ and $\Delta P_{\rm b}$, correlation functions $\rm Cor_X$ and $\rm Cor_P$, and average displacements $\bar{f}_k$ and $\bar{g}_k$ \cite{ber14,zho18,blu17}. Noting $\langle \Psi_{\rm g}|\hat{p}_k|\Psi_{\rm g}\rangle=0$, one has
\begin{eqnarray}
\label{phase var}
\Delta X_{\rm b} & = & \langle \Psi_{\rm g}|(\hat{x}_k)^2 |\Psi_{\rm g}\rangle - \langle \Psi_{\rm g}|\hat{x}_k|\Psi_{\rm g}\rangle^2, \nonumber \\
\Delta P_{\rm b} &= &\langle \Psi_{\rm g}|(\hat{p}_k)^2 |\Psi_{\rm g}\rangle,  \nonumber \\
{\rm Cor_X} & = & \langle \Psi_{\rm g}|\hat{x}_k \hat{x}_l |\Psi_{\rm g}\rangle -  \langle \Psi_{\rm g}|\hat{x}_k|\Psi_{\rm g}\rangle  \langle \Psi_{\rm g}|\hat{x}_l|\Psi_{\rm g}\rangle, \nonumber \\
{\rm Cor_P} & = & \langle \Psi_{\rm g}|\hat{p}_k \hat{p}_l |\Psi_{\rm g}\rangle,
\end{eqnarray}
where $\hat{x}_k$ and $\hat{p}_k$ represent quadrature operators for the phase-space variables, i.e., the position and momentum,
\begin{equation}
\label{vm_xp}
\hat{x}_k = \left(b_k+b_k^{\dag}\right)/\sqrt{2}, \qquad \hat{p}_k = i\left(b_k^{\dag}-b_k\right)/\sqrt{2},
\end{equation}
and the subscripts $k$ and $l$ correspond to the $k$-th and \break $l$-th bath modes, respectively.

To capture the characteristic of the ground-state wavefunction,  we introduce the average coherent-state weights
\begin{equation}
\label{vm_amp}
\overline A  =  \sqrt{\sum_{mn}A_mA_nF_{mn}}, \qquad \overline D  =  \sqrt{\sum_{mn}D_mD_nG_{mn}},
\end{equation}
and the average displacement coefficients
\begin{eqnarray}
\label{vm_dis}
\overline{f}_k & = & \sum_{m,n}\frac{A_mA_nF_{mn}(f_{m,k}+f_{n,k})}{2\overline{A}^2}, \nonumber \\
\overline{g}_k & = & \sum_{m,n}\frac{D_mD_nG_{mn}(g_{m,k}+g_{n,k})}{2\overline{D}^2},
\end{eqnarray}
where the functions $F_{mn}$ and $G_{mn}$ are defined as
\begin{eqnarray}
\label{vmfactor_2}
F_{mn} & = & \exp\left[-\frac{1}{2}\sum_{k}(f_{m,k}-f_{n,k})^2\right], \nonumber \\
G_{mn} & = & \exp\left[-\frac{1}{2}\sum_{k}(g_{m,k}-g_{n,k})^2\right].
\end{eqnarray}

Finally, the symmetries of the ground state are also probed here. In the case of $\varepsilon=0$ and $\Delta\neq 0$, the SBM may possess strong $\mathbb{Z}_2$ symmetry. Due to the competition between the tunneling and environmental dissipation, there exists a quantum phase transition separating a nondegenerate symmetric delocalized phase from a localized phase characterized by a doubly degenerate ground state. The projection operator from one branch to the other branch of the degenerate states is then introduced,
\begin{equation}
\label{project}
\hat{\mathcal {P}}=\sigma_x\exp{\left[i\pi \sum_{k=1}^{M}b_k^{\dag}b_k\right]}.
\end{equation}
The spontaneous breakdown of the $\mathbb{Z}_2$ symmetry can be described by the symmetry parameter defined as
\begin{equation}
\label{symmetry}
\zeta=\langle\Psi_g|\hat{\mathcal {P}}|\Psi_g\rangle \Delta_E,
\end{equation}
where $\Delta_E$ denotes a piecewise function of variable $E_g$, taking the values of $1$ if $E_g(\Psi_g)=E_g(\hat{\mathcal {P}}\Psi_g)$, and $0$ otherwise. Thereby the symmetry parameter is expected to be $\zeta=1$ ($\zeta=0$) for the delocalized (localized) phase, corresponding to the ground state with (without) the $\mathbb{Z}_2$ symmetry. In the biased case, i.e., $\varepsilon \neq 0$, the vanishing value of $\zeta$ holds for any coupling $\alpha$ since $\Delta_E=0$, indicating that the symmetry is always broken. Hence, $\zeta(\alpha)$ is a natural order parameter for quantum phase transitions associated with the spontaneous symmetry breaking.

\section{Numerical results}\label{sec:num}

The ground-state properties of the Ohmic SBM in a high dense spectrum are investigated with variational calculations in the weak tunneling limit, taking the setting of logarithmic discretization factor $\Lambda=1.01$ and tunneling amplitude $\Delta=0.01$ as an example. Theoretically, the number of effective bath modes $M \rightarrow \infty$ is required for the completeness of the environment. Considering the constraint available computational resources, a sufficiently large number $M=1000$ is used in main results. Besides, the spectral exponent $s=1$, number of coherent-superposition states $N=6$, and energy bias $\varepsilon=0$ are set unless noted otherwise. In numerical simulations, the statistical errors of the critical coupling and exponents are estimated by dividing the total samples into two subgroups. If the fluctuation in the frequency direction is comparable with or larger than the statistical error, it will be taken into account.

\subsection{Spontaneous symmetry breaking}
\begin{figure*}[ht]
\epsfysize=7cm \epsfclipoff \fboxsep=0pt
\setlength{\unitlength}{1.cm}
\begin{picture}(10,7)(0,0)
\put(-0.8, 0.0){{\epsffile{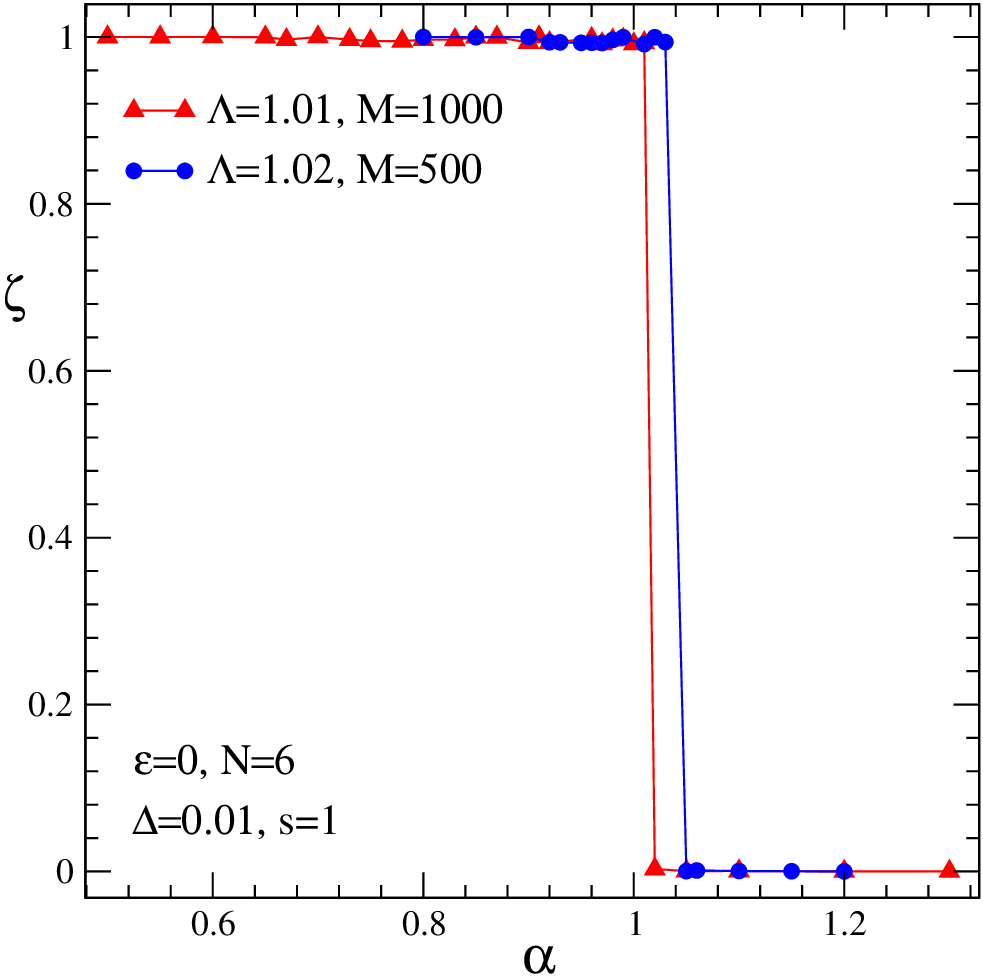}}}\epsfysize=7.2cm
\put(7.0, -0.2){{\epsffile{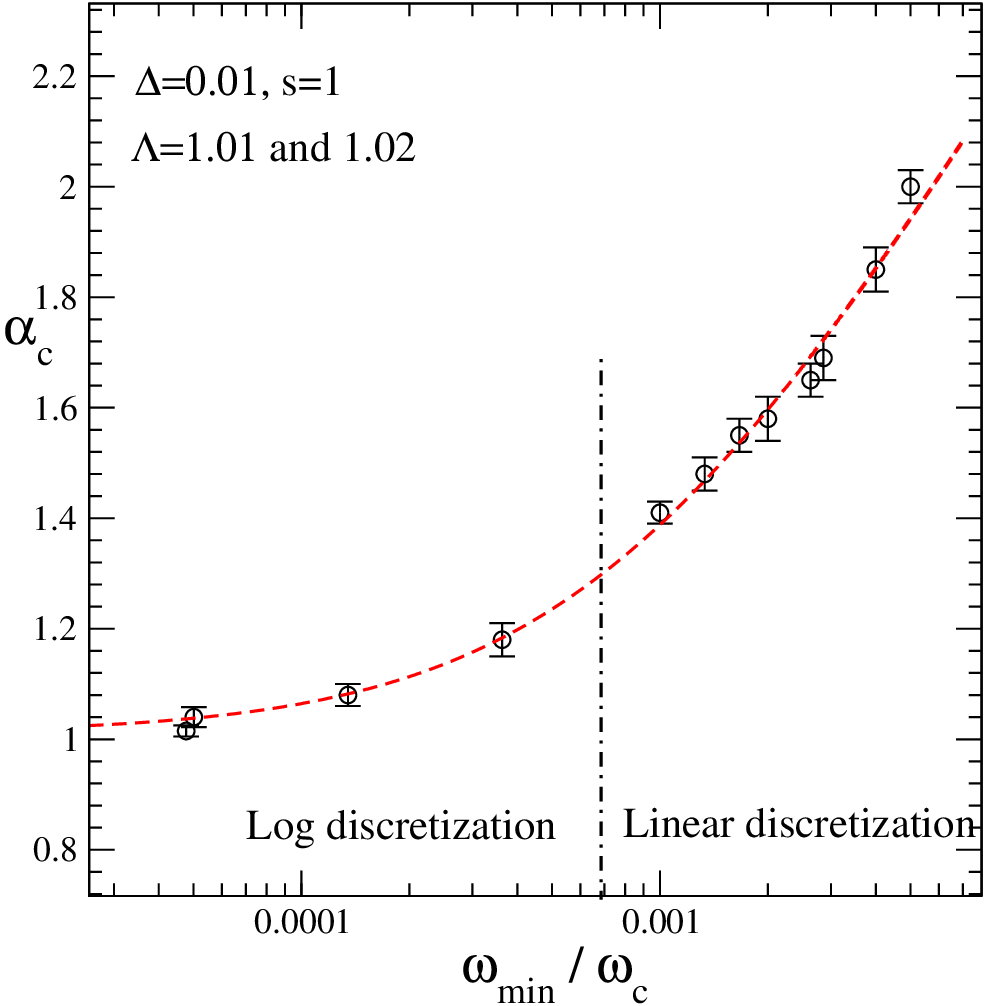}}}
\end{picture}

\hspace{2.0cm}\footnotesize{(a)}\hspace{7.0cm}\footnotesize{(b)}
\caption{(a) The symmetry parameter $\zeta$ defined in Eq.~(\ref{symmetry}) is plotted as a function of the coupling strength $\alpha$ on a linear scale. The tunneling amplitude $\Delta=0.01$ and logarithmic discretization factor $\Lambda=1.01$ and $1.02$ are used for the Ohmic SBM at $s=1$. (b) Displayed as a function of $\omega_{\rm min}/\omega_c$ on a linear-log scale is the transition boundary $\alpha_c$ obtained from the symmetry parameter $\zeta$. The results of the linear discretization are also given for $\omega_{\rm min}/\omega_c > 0.0007$. The dashed line represents the fit with a logarithmic form.
}
\label{f1}
\end{figure*}

Ground-state symmetries are firstly investigated with the symmetry parameter $\zeta$ defined in Eq.~(\ref{symmetry}). As shown in Fig.~\ref{f1}(a), $\zeta$ is displayed as a function of the coupling strength $\alpha$ for the logarithmic discretization factors $\Lambda=1.01$ and $1.02$ with the same low-energy cutoff $\omega_{\rm min}\approx 5\times 10^{-5}\omega_c$. The spontaneous symmetry breaking is confirmed by the emergence of the abrupt jump from $\zeta=1$ to $0$. The values of the critical point $\alpha_c=1.01(1)$ and $1.03(2)$ are then estimated, in agreement with $\alpha_c=1$. It indicates that the values of the logarithmic discretization factor $\Lambda$ are already sufficiently close to $1$ for the continuum limit $\Lambda \rightarrow 1$.

In Fig.~\ref{f1}(b), the transition boundary $\alpha_c$ is plotted against $\omega_{\rm min}/\omega_c$ on a linear-log scale. The results of the linear discretization are also presented for the lowest frequency $\omega_{\rm min}/\omega_c > 0.0007$ from supplementary calculations with $\omega_k = (k/M)\omega_c$. All of the data collapse onto a single curve, further confirming that the cases with $\Lambda=1.01$ and $1.02$ belong to the quasi linear discretization, yielding a high dense Ohmic spectrum. Using the fitting with the logarithmic form $y=a\ln(x+b)+c$, the asymptotic value $\alpha_c=1.0053$ is estimated by the extrapolation to $\omega_{\rm min}=0$, consistent with the renormalization group prediction $\alpha_c=1+\mathcal {O}(\Delta/ \omega_c)$ \cite{hur08}. By a linear dependence on the tunneling amplitude $\Delta$, one obtains the slope $(\alpha_c-1)\omega_c/\Delta=0.53$, in excellent agreement with the QMC one ($0.5$) estimated from $\alpha_c=1.05$ at $\Delta=0.1$ reported in Ref.~\cite{fil20}. Where the bath effects are taken into account by an effective Euclidean action whose kernel is expressed in terms of the continuous spectral density and bath propagator, instead of the discretization treatment of the Ohmic bath. Moreover, the prediction in this work for the frequency-range dependence of the critical coupling can be experimentally examined in the future.

\begin{figure*}[ht]
\centering
\epsfysize=8cm \epsfclipoff \fboxsep=0pt
\setlength{\unitlength}{1.cm}
\begin{picture}(9,8)(0,0)
\put(0.0,0.0){{\epsffile{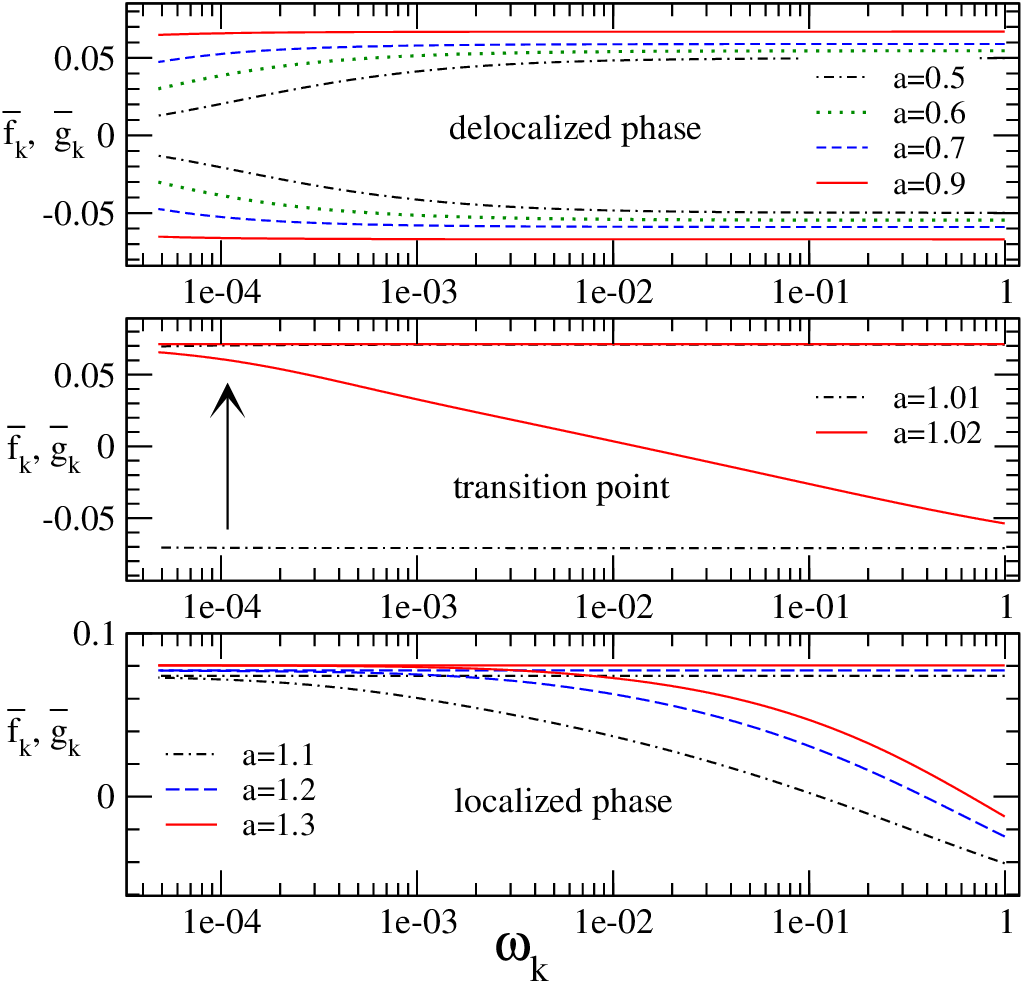}}}
\end{picture}
\caption{ The average displacement coefficients $\overline f_k$  and $\overline g_k$ for different couplings $\alpha$ are plotted with solid, dashed, dotted, and dash-dotted lines on a linear-log scale. Different behaviors are found in three panels from top to bottom, corresponding to the delocalized phase, transition point, and localized phase, respectively. The arrow indicates a huge jump of the average displacement coefficients in the low-frequency regime.
}
\label{f2}
\end{figure*}

For further understanding the symmetries, the average displacement coefficients $\overline f_k$ and $\overline g_k$ defined in Eq.~(\ref{vm_dis}) are measured at $\Lambda=1.01$ and $M=1000$ for different coupling strengths $\alpha$ and bath-mode frequencies $\omega_k$, as shown in Fig.~\ref{f2}. Taking $\alpha=0.5,0.6,0.7$ and $0.9$ as examples, a perfect antisymmetry relation $\overline f_k =- \overline g_k$ is observed over the whole range of frequencies $\omega_k$ in the upper panel, consistent with the usual assumption concerning the delocalized phase \cite{sil84,ber14}. For $\alpha=1.1,1.2$ and $1.3$, either $\overline f_k$ or $\overline g_k$ is equal to the classical displacement $\lambda_k/2\omega_k=\rm constant$, hence pointing to the localized phase. In the middle panel, a huge jump appears in the low-frequency asymptotic value of the displacement coefficient ($\overline f_k$ or $\overline g_k$) as the coupling strength $\alpha$ is changed by only a paltry amount of $0.01$. It again shows that the symmetry gets spontaneously broken at the critical coupling $\alpha_c=1.01(1)$.

\subsection{Quantum fluctuations and correlations}
\begin{figure*}[ht]
\epsfysize=7cm \epsfclipoff \fboxsep=0pt
\setlength{\unitlength}{1.cm}
\begin{picture}(10,7)(0,0)
\put(-1.2, -0.0){{\epsffile{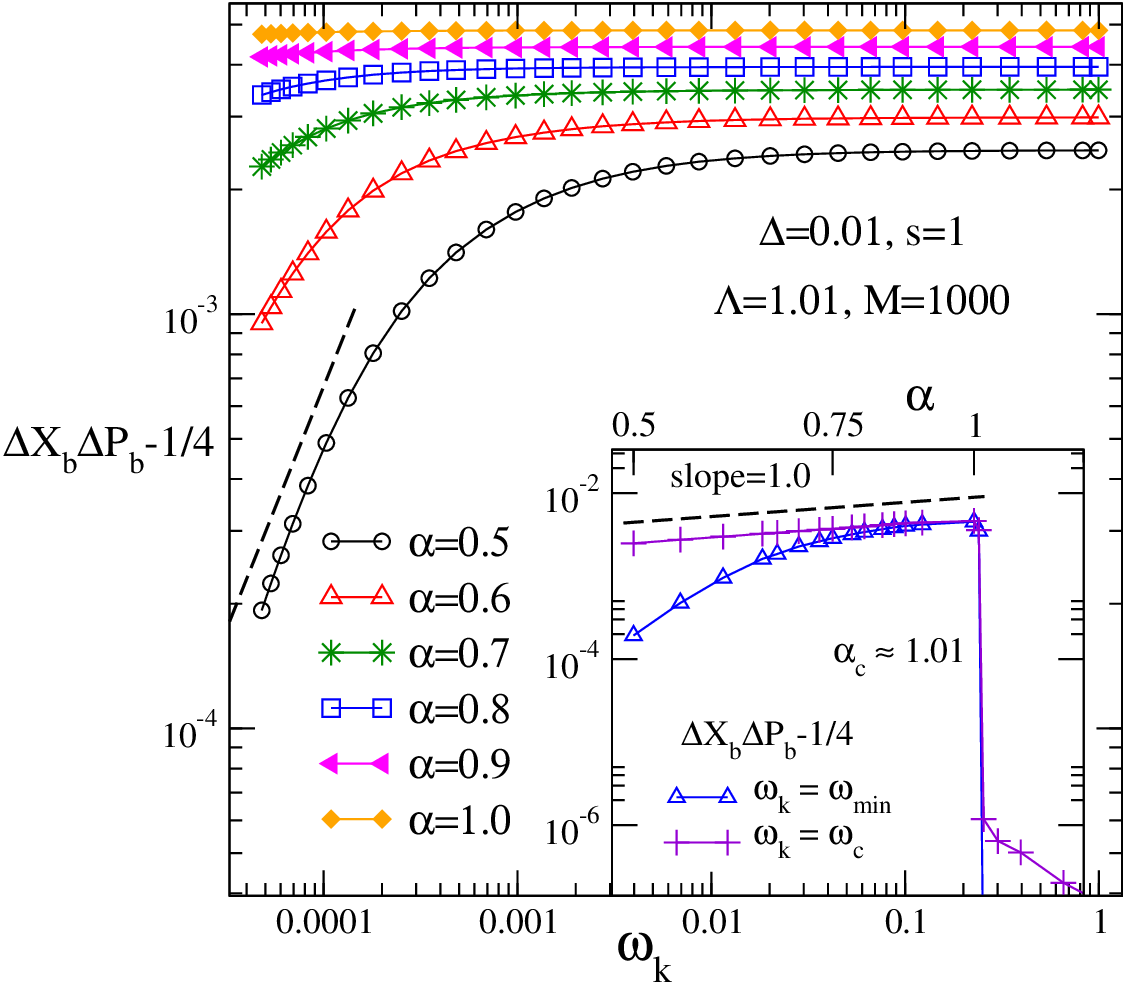}}}\epsfysize=7cm
\put(6.8, 0.0){{\epsffile{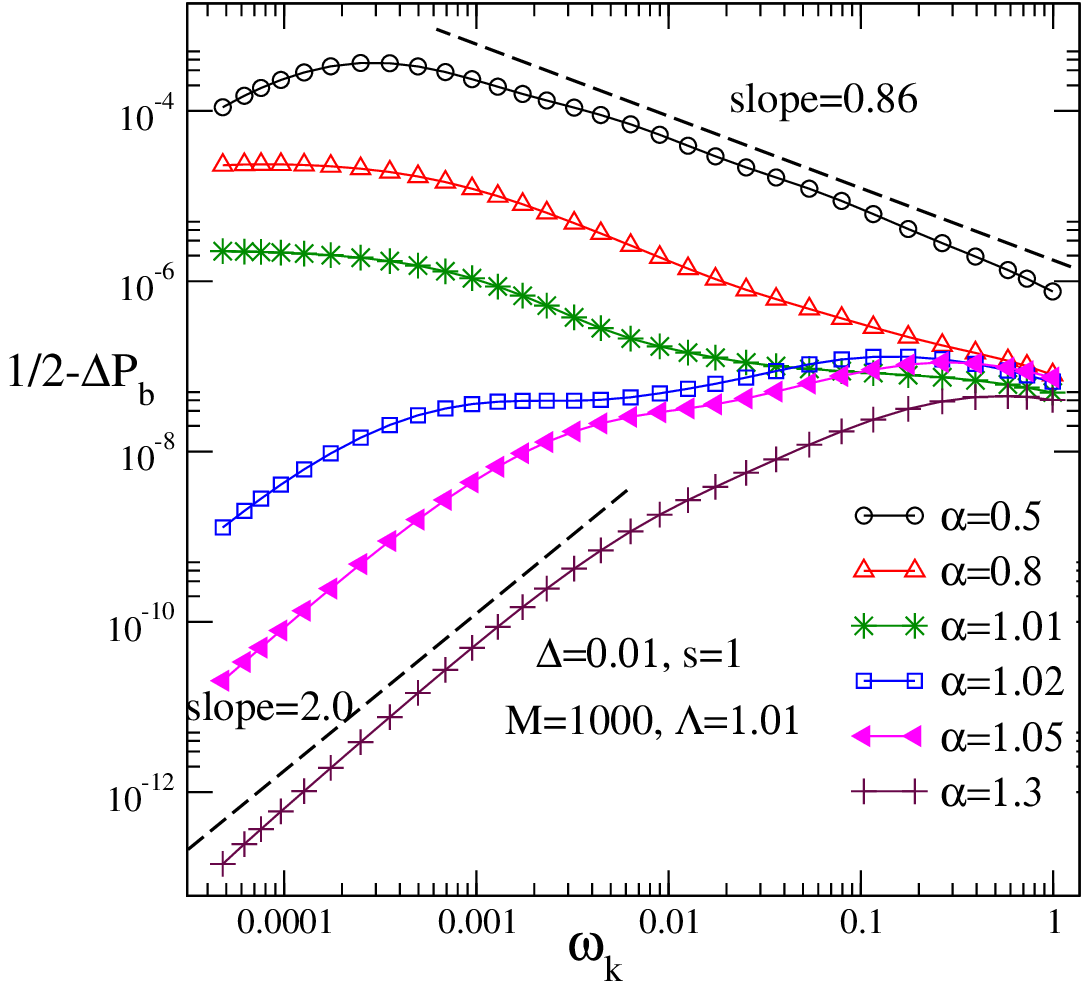}}}
\end{picture}

\hspace{2.5cm}\footnotesize{(a)}\hspace{7.5cm}\footnotesize{(b)}
\caption{(a) The departure from the minimum uncertainty, $\Delta X_{\rm b} \Delta P_{\rm b}- 1/4$, is shown for different coupling strengths $\alpha$ as a function of the frequency $\omega_k$ on a log-log scale. In the inset, the asymptotic values of $\Delta X_{\rm b} \Delta P_{\rm b}- 1/4$ are plotted in the low- and high-frequency limits. (b) The quantum fluctuation $1/2-\Delta P_{b}$ in the momentum space is displayed for different couplings $\alpha$. In both (a) and (b), dashed lines represent power-law fits. }
\label{f3}
\end{figure*}

In this subsection, quantum fluctuations and correlations in the Ohmic bath are investigated for the Kosterlitz-Thouless transition.  As single-coherent states obey minimum uncertainty relation $\Delta X_{\rm b}=\Delta P_{\rm b}=1/2$, quantum fluctuation from the coherent superposition in Eq.~(\ref{vmwave}) can be measured by the departure $\Delta X_{\rm b} \Delta P_{\rm b}- 1/4$. In Fig.~\ref{f3}(a), quantum fluctuation is plotted with respect to the frequency $\omega_k$ for various coupling strengths $\alpha$ on a log-log scale. It grows as a power law in the delocalized phase,  e.g., $\Delta X_{\rm b} \Delta P_{\rm b}- 1/4 \sim \omega_k^2$ at the Toulouse point $\alpha=0.5$, and gradually approaches to a $\alpha$-dependent constant value. Insets show the asymptotic values of the quantum fluctuations in the low- and high-frequency limits, taking the cases of $\omega_k =\omega_{\rm min}$ (solid line with open triangles) and $\omega_k =\omega_c$ (solid line with pluses) as examples. The transition point is located  at $\alpha_c = 1.01(1)$ by the drop of $\Delta X_{\rm b} \Delta P_{\rm b}- 1/4$ from $10^{-2}$ to $10^{-6}$. Moreover, the intersection of two curves suggests that the quantum fluctuation is independent of $\omega_k$ around the critical point $\alpha_c$. In the delocalized phase with $\alpha < \alpha_c$, a clean power-law behavior is found in the high-frequency limit, and the slope $1.0$ indicates that the saturation departure is proportional to the coupling. For the coupling $\alpha > \alpha_c$, the asymptotic values vanish in both two cases, confirming that the bath modes behave as a single-coherent state in the localized phase.

Quantum fluctuation of the momentum is also presented in Fig.~\ref{f3}(b) for different couplings $\alpha$ on a log-log scale. In contrast to $\Delta X_{\rm b} \Delta P_{\rm b}- 1/4$, the offset $1/2-\Delta P_{\rm b}$ in the delocalized phase shows a tendency to decay with the frequency $\omega_k$. Especially at the Toulouse point $\alpha=0.5$, a nice power-law decrease is found over more than three decades in frequencies, and the slope $\eta=0.86(1)$ is measured accurately. In the localized phase, the momentum fluctuation grows by more than four orders of magnitude, indicating that the value of $1/2-\Delta P_{\rm b}$ at the low frequency is negligibly small, as compared to those in the high-frequency region and in the delocalized phase. Besides, the slope $2.0$ is the same as that of $\Delta X_{\rm b} \Delta P_{\rm b}- 1/4$, suggesting that the power-law growth of quantum fluctuation is trivial in both two phases. Interestingly, a flattened curve can be inferred between $\alpha=1.01$ and $1.02$, pointing again that the quantum fluctuation is frequency-independent at the transition point.

In the recent work \cite{blu17}, two strong fingerprints of quantum criticality have been reported in the sub-Ohmic SBM.  One is an algebraic decay of the average displacement $\overline f_k \sim \omega_k^{(1-s)/2}$, and the other is a constant average squeezing amplitude which is related to the quantum fluctuation. In the Ohmic SBM with $s=1$, both the fingerprints are verified through our numerical work where $\overline f_k, \Delta X_{\rm b} \Delta P_{\rm b}- 1/4$, and $1/2-\Delta P_{\rm b}$ are frequency-independent at the transition point, as shown in Figs.~\ref{f2}, and \ref{f3}. In addition, a constant plateau of $\rm \Delta X_{\rm b} \Delta P_{\rm b}- 1/4$ is found in the delocalized phase $\alpha < \alpha_c$ for the frequencies $\omega_{k} \ge \omega^{*}$, corresponding to the critical domain. It confirms that the Ohmic bath possesses the quantum criticality even in the delocalized phase. It is quite similar with those in the low-temperature phase of the classical two-dimensional XY model, embodying the universality of the Kosterlitz-Thouless transition \cite{kos74}. Further analysis on the ground-state wave function gives that the above deviations from the minimum uncertainty relation $\Delta X_{\rm b}=\Delta P_{\rm b}=1/2$ are mainly caused by the effects of the antipolaron states which take place naturally in the delocalized phase \cite{ber14}.

\begin{figure*}[ht]
\epsfysize=7cm \epsfclipoff \fboxsep=0pt
\setlength{\unitlength}{1.cm}
\begin{picture}(10,7)(0,0)
\put(-0.8, 0.0){{\epsffile{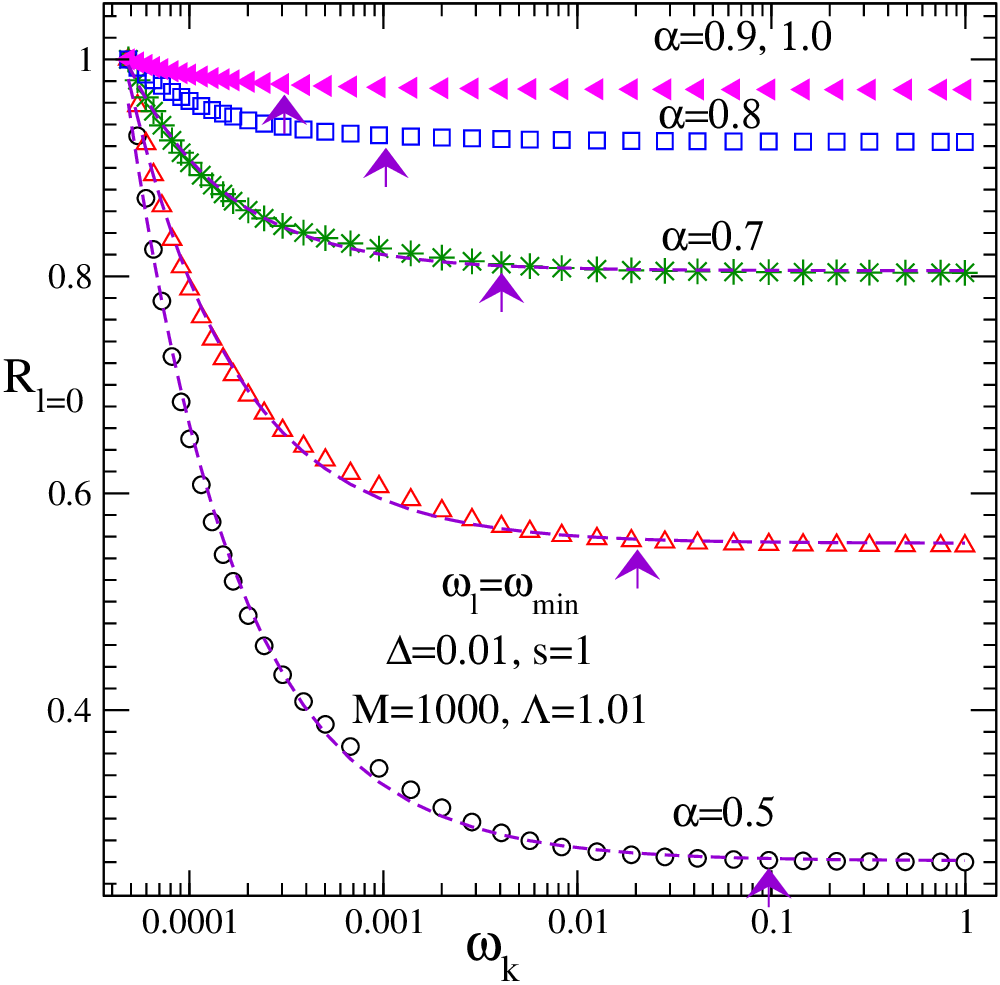}}}\epsfysize=7cm
\put(6.7, 0.0){{\epsffile{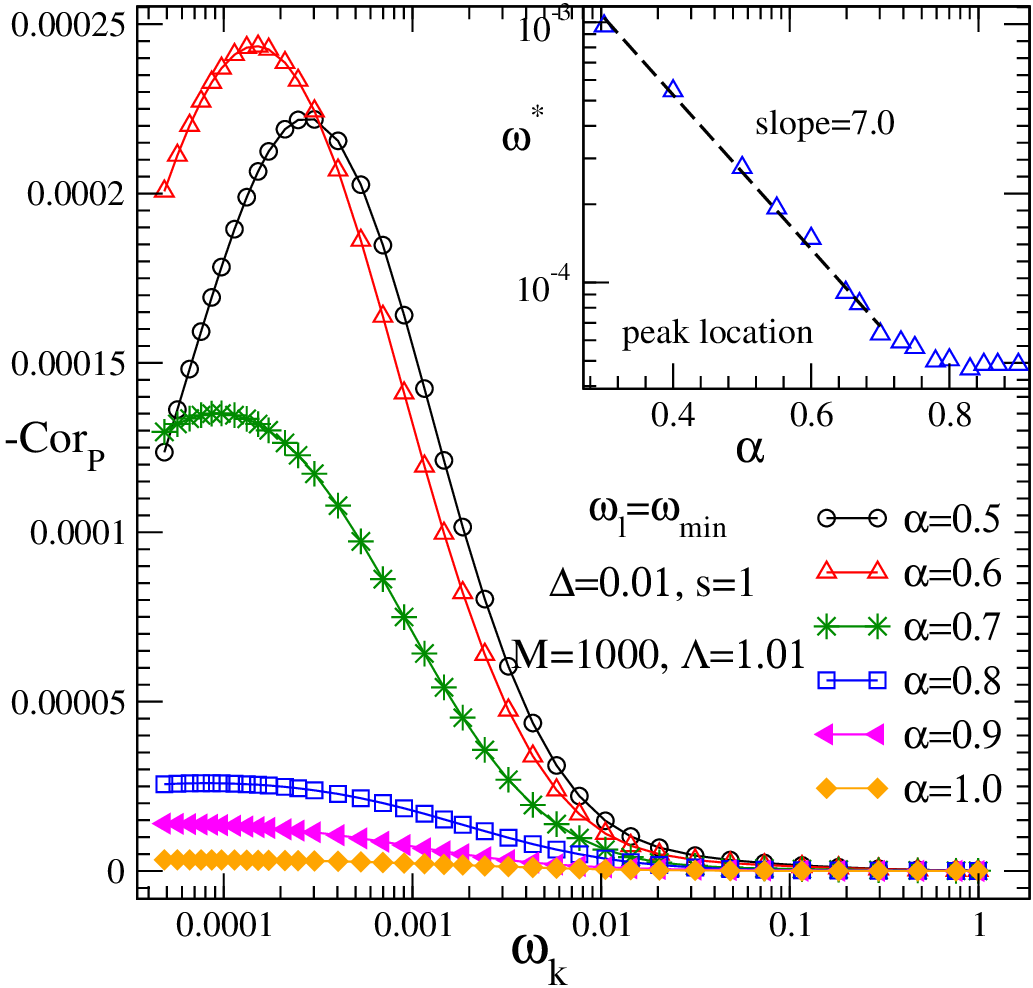}}}
\end{picture}

\hspace{2.0cm}\footnotesize{(a)}\hspace{7.2cm}\footnotesize{(b)}
\caption{(a) The correlation-fluctuation ratio function $R_{l=0}={\rm Cor}_X/(\Delta X_{\rm b}-1/2)$ at the fixed frequency  $\omega_l=\omega_{\rm min}$ is plotted  as a function of the frequency $\omega_k$ and coupling strength $\alpha$ on a linear-log scale. Other parameters $\Delta=0.01, s=1, \Lambda=1.01$, and $M=1000$ are set. Dashed lines show the best fits for the power-law decays. The transition frequencies $\omega^{*}$ beyond which the constant plateaus appear are marked by the arrows. (b) The $\omega_{k}$-dependent correlation function $\rm -Cor_{P}$ is plotted on a linear-log scale. Inset shows the optimal frequency $\omega^{*}$, and the dashed line represents an exponential fit.
}
\label{f4}
\end{figure*}

Besides quantum fluctuations, quantum correlations in the phase space, $\rm Cor_{X}$ and $\rm -Cor_{P}$ defined in Eq.~(\ref{phase var}), are also investigated as a function of the coupling $\alpha$ and two frequencies $\omega_{l}$ and $\omega_{k}$. Without loss of generality, the subscript $l=0$ is fixed for convenience, corresponding to $\omega_{l}=\omega_{\rm min}$. Similar with $\Delta X_{\rm b}\Delta P_{\rm b}-1/4$, quantum correlation $\rm Cor_{X}$ exhibits a smooth increase with the frequency $\omega_k$. It is in contrast to the general consensus on traditional statistical models, that is, the correlation function decaying with the distance. The possible reason is that all of the bath modes in SBM are uncoupled but simultaneously interact with the common spin system. To exclude the contribution of quantum fluctuation, we introduce the  correlation-fluctuation ratio function $R_{l=0}={\rm Cor}_X/(\Delta X_{\rm b}-1/2)$ instead.

As displayed in Fig.~\ref{f4}(a), the correlation-fluctuation ratio function $R_{l=0}$ decreases monotonically with increasing $\omega_k$, and approaches a $\alpha$-dependent constant. Dashed lines provide the power-law fitting to the numerical data, yielding the shift $\Delta R_{l=0} = R_{l=0}(\omega_k) - R_{l=0}(\omega_c) \sim \omega_k^{-\eta}$. The decaying exponent $\eta=0.85(2)$ at $\alpha=0.5$ agrees well with that in Fig.~\ref{f3}(b). Moreover, one observes the critical domain at high frequencies $\omega_k \ge \omega^*$, which gradually broadens into the whole frequency region as the coupling $\alpha$ increases, just the same as those of $\Delta X_{\rm b}\Delta P_{\rm b}-1/4$. It indicates the correlation length $\xi=1/\omega^*$ shows a tendency to diverge when $\alpha$ tends toward the critical coupling $\alpha_c=1$. An exponential increase of $\xi$ with the coupling $\alpha$ is then expected. In Fig.~\ref{f4}(b), the momentum correlation function $\rm -Cor_{P}$ exhibits bell-shaped relation, and the position of the peak decays with the coupling as $\omega^{*} \sim \exp(-7.0 \alpha)$ until it arrives at $\omega_{\rm min}$ when $\alpha > 0.8$,  consistent with the previous prediction. 

\begin{figure*}[ht]
\centering
\epsfysize=8cm \epsfclipoff \fboxsep=0pt
\setlength{\unitlength}{1.cm}
\begin{picture}(9,8)(0,0)
\put(0.0,0.0){{\epsffile{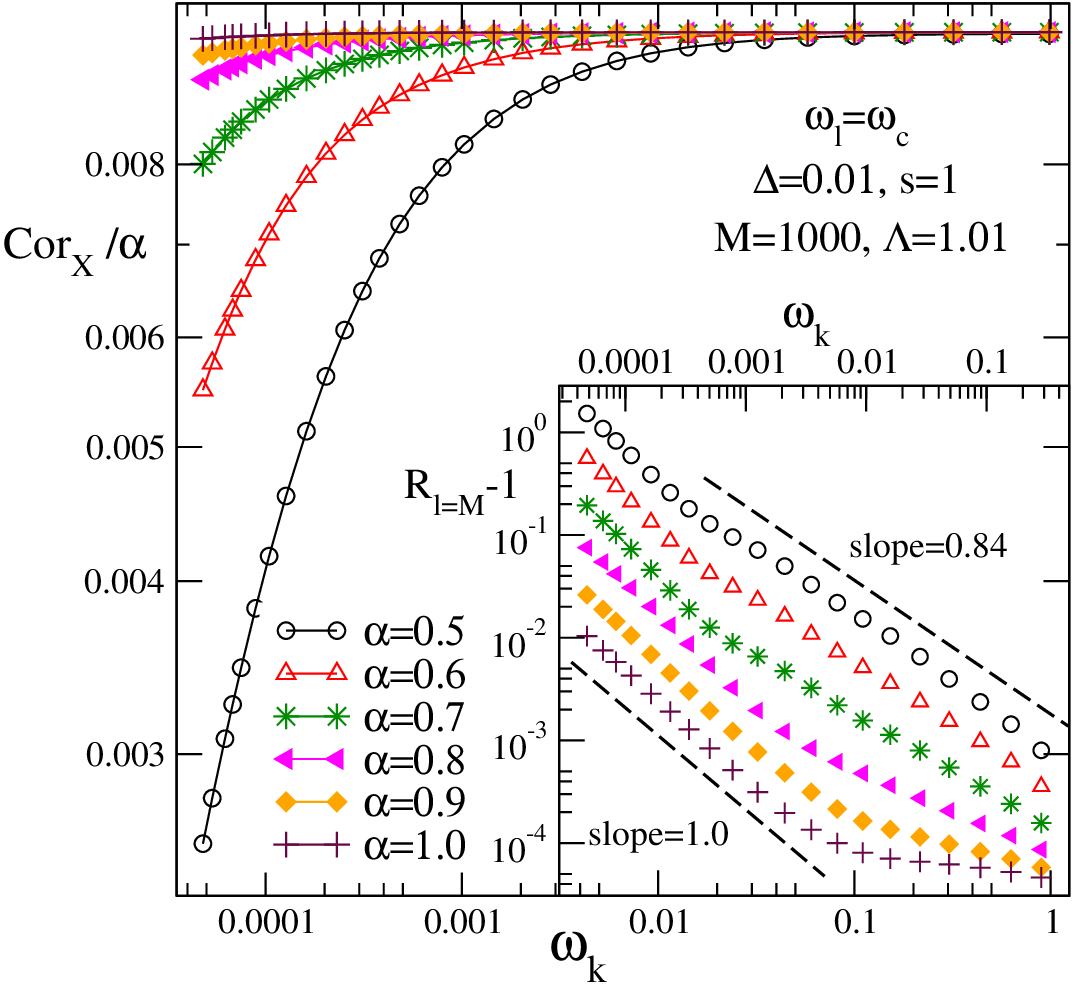}}}
\end{picture}
\caption{The scaled correlation function $\rm Cor_{X}/\alpha$ is plotted at the fixed frequency $\omega_l=\omega_c$  as a function of the bosonic frequency $\omega_k$ for different values of $\alpha$ on a log-log scale. Other parameters $\Delta=0.01, s=1, \Lambda=1.01$, and $M=1000$ are set. Inset shows the correlation-fluctuation ratio function $\rm R_{l=M}-1$ with respect to $\omega_k$ and $\alpha$. Dashed lines represent power-law fits.
}
\label{f5}
\end{figure*}

For comparison, quantum correlation $\rm Cor_{X}$ at another fixed frequency $\omega_l=\omega_c$ (i.e., $l=M$) is plotted in Fig.~\ref{f5}. For clarity, it is rescaled by a factor $1/\alpha$. One clearly observes that the curves of different $\alpha$ overlap at high frequencies, confirming the linear coupling dependence of $\rm Cor_{X}$, the same as that of $\Delta X_{\rm b} \Delta P_{\rm b}- 1/4$. Since the correlation-fluctuation ratio is $R_{l=M} \equiv 1$ at the cutoff frequency $\omega_c$, inset shows the offset $\rm R_{l=M}-1$ as a function of $\omega_k$ for different coupling $\alpha$ on a log-log scale. It exhibits a power-law decay at the Toulouse point $\alpha=0.5$, and the slope $\eta=0.84(2)$ is again consistent with the one in Fig.~\ref{f3}(b). For the coupling strength close to the transition point, e.g., $\alpha=1.0$, the function $R_{l=M}(\omega_k)-1$ exhibits a power-law behavior, too, and the decay is a litter faster than that at $\alpha=0.5$.

\subsection{Validity of variational calculations}
\begin{figure*}[htb]
\epsfysize=7cm \epsfclipoff \fboxsep=0pt
\setlength{\unitlength}{1.cm}
\begin{picture}(10,7)(0,0)
\put(-1.0, 0.0){{\epsffile{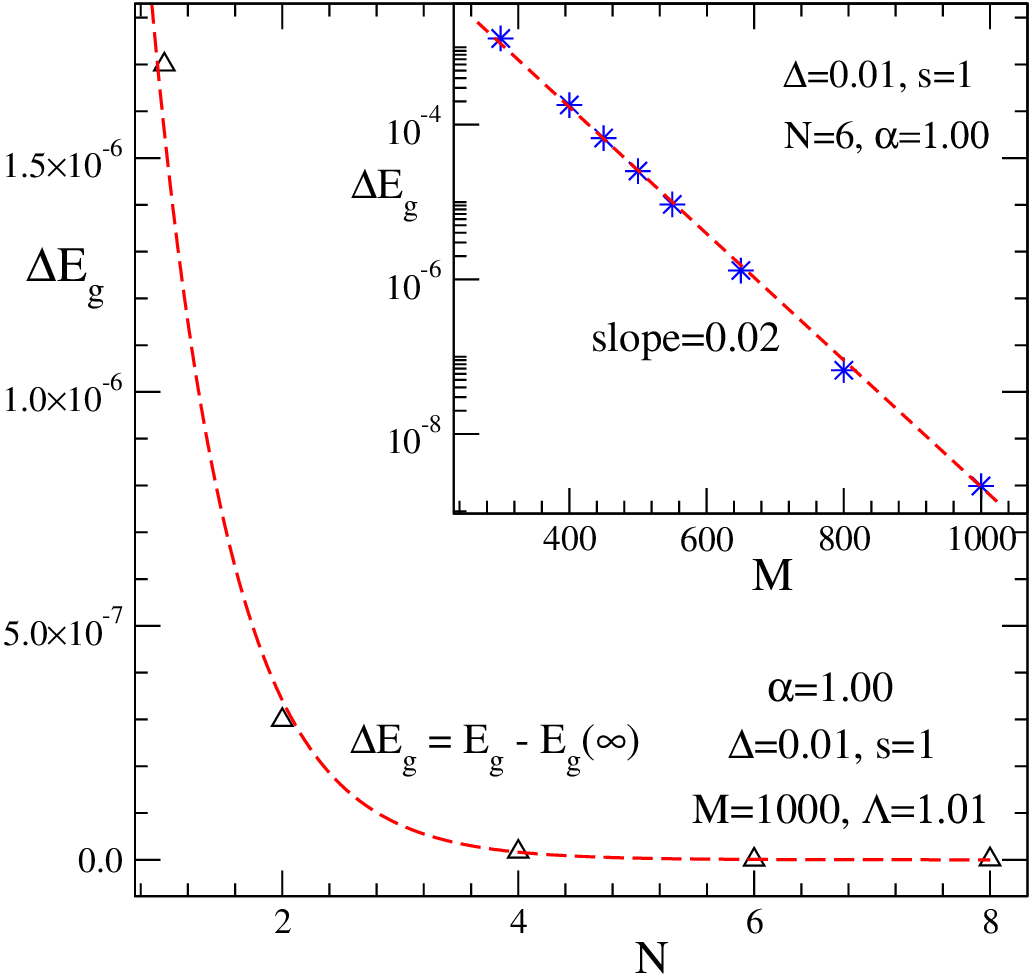}}}\epsfysize=7cm
\put(7.0, 0.0){{\epsffile{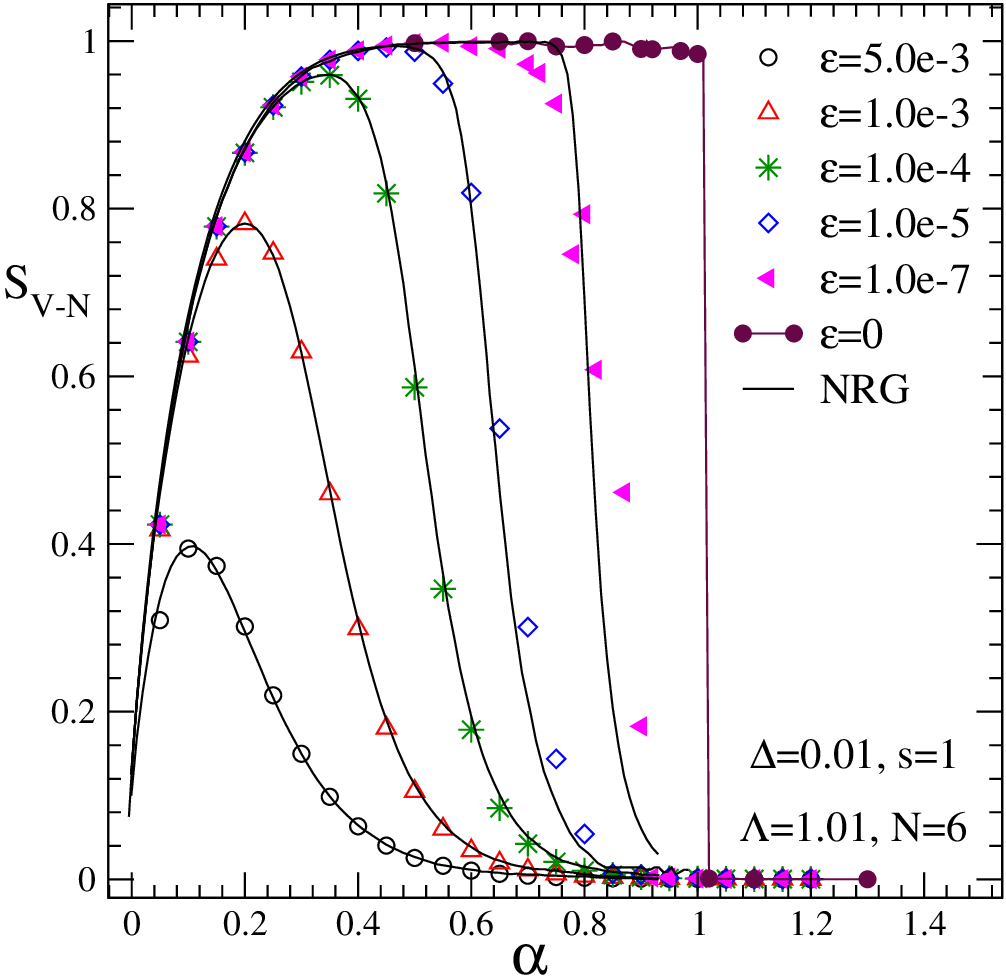}}}
\end{picture}

\hspace{2.0cm}\footnotesize{(a)}\hspace{7.2cm}\footnotesize{(b)}
\caption{(a) The convergence of ground-state energy $E_g$ is displayed with respective to the numbers of the coherent superposition states
$N$ and effective bath modes $M$ (in the inset) in the case of $s = 1, \alpha = 1, \Delta = 0.01$, and $\Lambda=1.01$. Dashed lines show exponential fits to the ground-state energy shift $\Delta E_g=E_g - E_g(\infty)$. (b) The von Neumann entropy $S_{\rm v-N}$ is plotted as a function of $\alpha$ for different values of the bias $\varepsilon =0.005,10^{-3},10^{-4},10^{-5},10^{-7}$ and $0$ (from left to right). For comparison, the results of NRG calculations are also shown with solid lines.
}
\label{f6}
\end{figure*}

The validity of the variational approach is carefully examined in this subsection. Firstly, the convergency test of the ground-state energy $E_g$ is performed with respect to the numbers of the coherent superposition states $N$ and effective bath modes $M$ defined in Eq.~(\ref{vmwave}), taking the case of $\alpha = 1, \Delta = 0.01$, and $\Lambda=1.01$ as an example. In Fig.\ref{f6}(a), the energy shift $\Delta E_g=E_g(N) - E_g(\infty)$ is shown for a fixed parameter $M=1000$ on a linear scale, where $E_g(\infty)$ is the asymptotic value of the ground-state energy. As $\alpha$ increases, the shift decays exponentially as $\Delta E_g \sim \exp(-1.5N)$. The significantly large slope suggests that a small value of $N$, i.e., $N=6$, is sufficient to study the ground-state phase transitions of Ohmic SBM via the variational approach. Moreover, the dependence of $\Delta E_g$ on the bath-mode number $M$ is demonstrated in the inset of Fig.\ref{f6}(a) on a linear-log scale at $N=6$. Similarly, an exponential decay of $\Delta E_g$ is observed with the slope $0.02$, showing that $M=1000$ is sufficient for the convergence.

Subsequently, extension to the biased Ohmic SBM is performed for the spin-related observations, such as the spin magnetization $\langle\sigma_z\rangle $, spin coherence $\langle\sigma_x\rangle $, and von-Neumann entropy $S_{\rm v-N}$ which denotes the entanglement between the spin and surrounding bath, $S_{\rm{v-N}}=-\omega_{+}\log\omega_{+}-\omega_{-}\log\omega_{-}$ where $\omega_{\pm}= (1\pm\sqrt{\langle{\sigma_x}\rangle^2+\langle{\sigma_y}\rangle^2+\langle{\sigma_z}\rangle^2})/2$.
For comparison, the results of NRG calculations are also given with the parameters, e.g., logarithmic discretization factor $\Lambda=2$, lowest energy levels $N_{s}= 150$, and bosonic truncated number $N_b=8$, the same as those in the earlier work~\cite{hur07}.

Taking the von-Neumann entropy $S_{\rm v-N}$ presented in Fig.~\ref{f6}(b) as a representative example, the results of NVM and NRG approaches agree well for the biases $\varepsilon=0.005,10^{-3},10^{-4}$, and $10^{-5}$, although there is a slight deviation under a weaker bias $\varepsilon=1.0\times10^{-7}$. It indicates that both of these two approaches are available to obtain an accurate description of the ground state. In addition, an infinitesimal but nonvanishing bias is usually used in NRG calculations to lift the degeneracy \cite{ort10}. Even under a tiny bias $\varepsilon=1.0 \times10^{-7}$, however, a sharp crossover occurs instead of the discontinuity, and the transition point estimated from the abrupt jump of NRG curve is obviously smaller than $\alpha_c=1$, as shown in the subfigure. In contrast, the value of $\alpha_c=1.01(1)$ from NVM calculations with the vanishing bias $\varepsilon=0$ is consistent with the theoretical prediction $\alpha_c \rightarrow 1^{+}$, thereby lending support to the superiority of the variational calculations.

\section{Conclusion}\label{sec:con}

By performing large-scale numerical variational calculations with a quasi-linear discretization, we have presented a comprehensive study of the ground-state quantum phase transitions of Ohmic SBM in a high dense spectrum in the weak tunneling limit, using the bare tunneling amplitude $\Delta=0.01$ and discretization factor $\Lambda=1.01$. The asymptotic value of the critical coupling $\alpha_c=1.0053$ has been accurately determined by extrapolation to $\omega_{\rm min}=0$, in good agreement with the theoretical prediction $\alpha_c=1+\mathcal {O}(\Delta/ \omega_c)$ \cite{hur08} and very recent numerical results obtained by the imaginary-time propagation \cite{wan19,fil20}. The values of the exponent $\eta=0.85(2)$ and $0$ have been measured from the quantum fluctuations and correlations in the Ohmic bath at the Toulouse point $\alpha=0.5$ and transition point $\alpha_c$, respectively. In addition, quantum criticality of Ohmic bath has been demonstrated explicitly both in the delocalized phase and at the transition point, lending support to the quantum phase transition of the Kosterlitz-Thouless type.

Very recently, quantum simulations of the spin-boson model have been realized by using a superconducting qubit connected to a microwave circuit wherein the tunability of the interaction allows one to observe quantum phase transitions \cite{lep18,yam19}. Our work provides the prediction on the $\omega_{\rm min}$ dependence of the transition point which can been experimental examined in the future, and the guidance for the choices of the Ohmic-bath frequency range $\omega_c/\omega_{\rm min}$ in experiments to achieve the exact transition point $\alpha_c=1$.

{\bf Acknowledgements:} This work was supported in part by National Natural Science Foundation of China under Grant Nos. $11875120$.

\section*{References}
\bibliography{sbm}

\begin{thebibliography}{10}
\expandafter\ifx\csname url\endcsname\relax
  \def\url#1{\texttt{#1}}\fi
\expandafter\ifx\csname urlprefix\endcsname\relax\def\urlprefix{URL }\fi
\expandafter\ifx\csname href\endcsname\relax
  \def\href#1#2{#2} \def\path#1{#1}\fi

\bibitem{wei07}
U.~Weiss, Quantum Dissipative Systems, 3rd Edition, World Scientific,
  Singapore, 2007.

\bibitem{hur10}
K.~L. Hur, CRC Press, Boca Raton, 2010, Ch.~9, pp. 217--240.

\bibitem{sac11}
S.~Sachdev, Quantum Phase Transitions, 2nd Edition, Cambridge University Press,
  Cambridge, England, 2011.

\bibitem{leg87}
A.~J. Leggett, S.~Chakravarty, A.~T. Dorsey, M.~P.~A. Fisher, A.~Garg,
  W.~Zwerger, Dynamics of the dissipative two-state system, Rev. Mod. Phys. 59
  (1987) 1--85.
\newblock \href {http://dx.doi.org/10.1103/RevModPhys.59.1}
  {\path{doi:10.1103/RevModPhys.59.1}}.

\bibitem{hur08}
K.~L. Hur, Entanglement entropy, decoherence, and quantum phase transitions of
  a dissipative two-level system, Annals of Physics 323 (2008) 2208 -- 2240.
\newblock \href {http://dx.doi.org/10.1016/j.aop.2007.12.003}
  {\path{doi:10.1016/j.aop.2007.12.003}}.

\bibitem{bre16}
H.~P. Breuer, E.~M. Laine, J.~Piilo, B.~Vacchini, Colloquium: Non-markovian
  dynamics in open quantum systems, Rev. Mod. Phys. 88 (2016) 021002.
\newblock \href {http://dx.doi.org/10.1103/RevModPhys.88.021002}
  {\path{doi:10.1103/RevModPhys.88.021002}}.

\bibitem{lew88}
J.~T. Lewis, G.~A. Raggio, The equilibrium thermodynamics of a spin-boson
  model, J. Stat. Phys 50 (1988) 1201.
\newblock \href {http://dx.doi.org/10.1007/BF01019161}
  {\path{doi:10.1007/BF01019161}}.

\bibitem{gol92}
B.~Golding, N.~M. Zimmerman, S.~N. Coppersmith, Dissipative quantum tunneling
  of a single microscopic defect in a mesoscopic metal, Phys. Rev. Lett. 68
  (1992) 998--1001.
\newblock \href {http://dx.doi.org/10.1103/PhysRevLett.68.998}
  {\path{doi:10.1103/PhysRevLett.68.998}}.

\bibitem{cha95}
S.~Chakravarty, J.~Rudnick, Dissipative dynamics of a two-state system, the
  kondo problem, and the inverse-square ising model, Phys. Rev. Lett. 75 (1995)
  501--504.
\newblock \href {http://dx.doi.org/10.1103/PhysRevLett.75.501}
  {\path{doi:10.1103/PhysRevLett.75.501}}.

\bibitem{eng07}
G.~S. Engel, T.~R. Calhoun, E.~L. Read, T.~K. Ahn, T.~Mancal, Y.~C. Cheng,
  R.~E. Blankenship, G.~R. Fleming, Evidence for wavelike energy transfer
  through quantum coherence in photosynthetic systems, Nature 446 (2007)
  782--786.
\newblock \href {http://dx.doi.org/10.1038/nature05678}
  {\path{doi:10.1038/nature05678}}.

\bibitem{col09}
E.~Collini, G.~D. Scholes, Coherent intrachain energy migration in a conjugated
  polymer at room temperature, Science 323 (2009) 369--373.
\newblock \href {http://dx.doi.org/10.1126/science.1164016}
  {\path{doi:10.1126/science.1164016}}.

\bibitem{gar17}
L.~Garbe, I.~L. Egusquiza, E.~Solano, C.~Ciuti, T.~Coudreau, P.~Milman,
  S.~Felicetti, Superradiant phase transition in the ultrastrong-coupling
  regime of the two-photon dicke model, Phys. Rev. A 95 (2017) 053854.
\newblock \href {http://dx.doi.org/10.1103/PhysRevA.95.053854}
  {\path{doi:10.1103/PhysRevA.95.053854}}.

\bibitem{ota11}
Y.~Ota, S.~Iwamoto, N.~Kumagai, Y.~Arakawa, Spontaneous two-photon emission
  from a single quantum dot, Phys. Rev. Lett. 107 (2011) 233602.
\newblock \href {http://dx.doi.org/10.1103/PhysRevLett.107.233602}
  {\path{doi:10.1103/PhysRevLett.107.233602}}.

\bibitem{por08}
D.~Porras, F.~Marquardt, J.~von Delft, J.~I. Cirac, Mesoscopic spin-boson
  models of trapped ions, Phys. Rev. A 78 (2008) 010101(R).
\newblock \href {http://dx.doi.org/10.1103/PhysRevA.78.010101}
  {\path{doi:10.1103/PhysRevA.78.010101}}.

\bibitem{uzd15}
R.~Uzdin, A.~Levy, R.~Kosloff, Equivalence of quantum heat machines, and
  quantum-thermodynamic signatures, Phys. Rev. X 5 (2015) 031044.
\newblock \href {http://dx.doi.org/10.1103/PhysRevX.5.031044}
  {\path{doi:10.1103/PhysRevX.5.031044}}.

\bibitem{lep18}
J.~Lepp\"akangas, J.~Braum\"uller, M.~Hauck, J.-M. Reiner, I.~Schwenk,
  S.~Zanker, L.~Fritz, A.~V. Ustinov, M.~Weides, M.~Marthaler, Quantum
  simulation of the spin-boson model with a microwave circuit, Phys. Rev. A 97
  (2018) 052321.
\newblock \href {http://dx.doi.org/10.1103/PhysRevA.97.052321}
  {\path{doi:10.1103/PhysRevA.97.052321}}.

\bibitem{sil84}
R.~Silbey, R.~A. Harris, Variational calculation of the dynamics of a two level
  system interacting with a bath, J. Chem. Phys. 80 (1984) 2615--2617.
\newblock \href {http://dx.doi.org/10.1063/1.447055}
  {\path{doi:10.1063/1.447055}}.

\bibitem{chi11}
A.~W. Chin, J.~Prior, S.~F. Huelga, M.~B. Plenio, Generalized polaron ansatz
  for the ground state of the sub-ohmic spin-boson model: An analytic theory of
  the localization transition, Phys. Rev. Lett. 107 (2011) 160601.
\newblock \href {http://dx.doi.org/10.1103/PhysRevLett.107.160601}
  {\path{doi:10.1103/PhysRevLett.107.160601}}.

\bibitem{naz12}
A.~Nazir, D.~P.~S. McCutcheon, A.~W. Chin, Ground state and dynamics of the
  biased dissipative two-state system: Beyond variational polaron theory, Phys.
  Rev. B 85 (2012) 224301.
\newblock \href {http://dx.doi.org/10.1103/PhysRevB.85.224301}
  {\path{doi:10.1103/PhysRevB.85.224301}}.

\bibitem{ber14}
S.~Bera, A.~Nazir, A.~W. Chin, H.~U. Baranger, S.~Florens, Generalized
  multipolaron expansion for the spin-boson model: Environmental entanglement
  and the biased two-state system, Phys. Rev. B 90 (2014) 075110.
\newblock \href {http://dx.doi.org/10.1103/PhysRevB.90.075110}
  {\path{doi:10.1103/PhysRevB.90.075110}}.

\bibitem{wu17}
W.~Wu, J.~B. Xu, Quantum coherence of spin-boson model at finite temperature,
  Annals of Physics 377 (2017) 48 -- 61.
\newblock \href {http://dx.doi.org/https://doi.org/10.1016/j.aop.2017.01.014}
  {\path{doi:https://doi.org/10.1016/j.aop.2017.01.014}}.

\bibitem{pin18}
M.~Pino, J.~J. Garc{\'{\i}}a-Ripoll, Quantum annealing in spin-boson model:
  from a perturbative to an ultrastrong mediated coupling, New Journal of
  Physics 20 (2018) 113027.
\newblock \href {http://dx.doi.org/10.1088/1367-2630/aaeeea}
  {\path{doi:10.1088/1367-2630/aaeeea}}.

\bibitem{nal13}
P.~Nalbach, M.~Thorwart, Crossover from coherent to incoherent quantum dynamics
  due to sub-ohmic dephasing, Phys. Rev. B 87 (2013) 014116.
\newblock \href {http://dx.doi.org/10.1103/PhysRevB.87.014116}
  {\path{doi:10.1103/PhysRevB.87.014116}}.

\bibitem{guo12}
C.~Guo, A.~Weichselbaum, J.~von Delft, M.~Vojta, Critical and strong-coupling
  phases in one- and two-bath spin-boson models, Phys. Rev. Lett. 108 (2012)
  160401.
\newblock \href {http://dx.doi.org/10.1103/PhysRevLett.108.160401}
  {\path{doi:10.1103/PhysRevLett.108.160401}}.

\bibitem{zhou14}
N.~J. Zhou, L.~P. Chen, Y.~Zhao, D.~Mozyrsky, V.~Chernyak, Y.~Zhao,
  Ground-state properties of sub-ohmic spin-boson model with simultaneous
  diagonal and off-diagonal coupling, Phys. Rev. B 90 (2014) 155135.
\newblock \href {http://dx.doi.org/10.1103/PhysRevB.90.155135}
  {\path{doi:10.1103/PhysRevB.90.155135}}.

\bibitem{zho18}
N.~J. Zhou, Y.~Y. Zhang, Z.~G. L\"u, Y.~Zhao, Variational study of the
  two-impurity spin-boson model with a common ohmic bath: Ground-state phase
  transitions, Annalen der Physik 530 (2018) 1800120.
\newblock \href {http://dx.doi.org/10.1002/andp.201800120}
  {\path{doi:10.1002/andp.201800120}}.

\bibitem{wan20}
Y.~Z. Wang, S.~He, L.~W. Duan, Q.~H. Chen, Rich phase diagram of quantum phases
  in the anisotropic subohmic spin-boson model, Phys. Rev. B 101 (2020) 155147.
\newblock \href {http://dx.doi.org/10.1103/PhysRevB.101.155147}
  {\path{doi:10.1103/PhysRevB.101.155147}}.

\bibitem{gui85}
F.~Guinea, V.~Hakim, A.~Muramatsu, Bosonization of a two-level system with
  dissipation, Phys. Rev. B 32 (1985) 4410--4418.
\newblock \href {http://dx.doi.org/10.1103/PhysRevB.32.4410}
  {\path{doi:10.1103/PhysRevB.32.4410}}.

\bibitem{ort10}
P.~P. Orth, D.~Roosen, W.~Hofstetter, K.~Le~Hur, Dynamics, synchronization, and
  quantum phase transitions of two dissipative spins, Phys. Rev. B 82 (2010)
  144423.
\newblock \href {http://dx.doi.org/10.1103/PhysRevB.82.144423}
  {\path{doi:10.1103/PhysRevB.82.144423}}.

\bibitem{mcc10}
D.~P.~S. McCutcheon, A.~Nazir, S.~Bose, A.~J. Fisher, Separation-dependent
  localization in a two-impurity spin-boson model, Phys. Rev. B 81 (2010)
  235321.
\newblock \href {http://dx.doi.org/10.1103/PhysRevB.81.235321}
  {\path{doi:10.1103/PhysRevB.81.235321}}.

\bibitem{win14}
A.~Winter, H.~Rieger, Quantum phase transition and correlations in the
  multi-spin-boson model, Phys. Rev. B 90 (2014) 224401.
\newblock \href {http://dx.doi.org/10.1103/PhysRevB.90.224401}
  {\path{doi:10.1103/PhysRevB.90.224401}}.

\bibitem{mag18}
L.~Magazz\`u, P.~Forn-D\'iaz, R.~Belyansky, J.-L. Orgiazzi, M.~A. Yurtalan,
  M.~R. Otto, A.~Lupascu, C.~M. Wilson, M.~Grifoni, Probing the strongly driven
  spin-boson model in a superconducting quantum circuit, Nature Communications
  9 (2018) 1403.
\newblock \href {http://dx.doi.org/10.1038/s41467-018-03626-w}
  {\path{doi:10.1038/s41467-018-03626-w}}.

\bibitem{voj05}
M.~Vojta, N.~H. Tong, R.~Bulla, Quantum phase transitions in the sub-ohmic
  spin-boson model: Failure of the quantum-classical mapping, Phys. Rev. Lett.
  94 (2005) 070604.
\newblock \href {http://dx.doi.org/10.1103/PhysRevLett.94.070604}
  {\path{doi:10.1103/PhysRevLett.94.070604}}.

\bibitem{alv09}
A.~Alvermann, H.~Fehske, Sparse polynomial space approach to dissipative
  quantum systems: Application to the sub-ohmic spin-boson model, Phys. Rev.
  Lett. 102 (2009) 150601.
\newblock \href {http://dx.doi.org/10.1103/PhysRevLett.102.150601}
  {\path{doi:10.1103/PhysRevLett.102.150601}}.

\bibitem{win09}
A.~Winter, H.~Rieger, M.~Vojta, R.~Bulla, Quantum phase transition in the
  sub-ohmic spin-boson model: Quantum monte carlo study with a continuous
  imaginary time cluster algorithm, Phys. Rev. Lett. 102 (2009) 030601.
\newblock \href {http://dx.doi.org/10.1103/PhysRevLett.102.030601}
  {\path{doi:10.1103/PhysRevLett.102.030601}}.

\bibitem{zha10}
Y.~Y. Zhang, Q.~H. Chen, K.~L. Wang, Quantum phase transition in the sub-ohmic
  spin-boson model: An extended coherent-state approach, Phys. Rev. B 81 (2010)
  121105(R).
\newblock \href {http://dx.doi.org/10.1103/PhysRevB.81.121105}
  {\path{doi:10.1103/PhysRevB.81.121105}}.

\bibitem{bul05}
R.~Bulla, H.~J. Lee, N.~H. Tong, M.~Vojta, Numerical renormalization group for
  quantum impurities in a bosonic bath, Phys. Rev. B 71 (2005) 045122.
\newblock \href {http://dx.doi.org/10.1103/PhysRevB.71.045122}
  {\path{doi:10.1103/PhysRevB.71.045122}}.

\bibitem{wan19}
H.~Wang, J.~Shao, Quantum phase transition in the spin-boson model: A
  multilayer multiconfiguration time-dependent hartree study, J. Phys. Chem. A
  123 (2019) 1882--1893.
\newblock \href {http://dx.doi.org/10.1021/acs.jpca.8b11136}
  {\path{doi:10.1021/acs.jpca.8b11136}}.

\bibitem{fil20}
G.~De~Filippis, A.~de~Candia, L.~M. Cangemi, M.~Sassetti, R.~Fazio,
  V.~Cataudella, Quantum phase transitions in the spin-boson model: Monte carlo
  method versus variational approach \`a la feynman, Phys. Rev. B 101 (2020)
  180408(R).
\newblock \href {http://dx.doi.org/10.1103/PhysRevB.101.180408}
  {\path{doi:10.1103/PhysRevB.101.180408}}.

\bibitem{zhe15}
H.~Zheng, Z.~G. L\"u, Y.~Zhao, Ansatz for the quantum phase transition in a
  dissipative two-qubit system, Phys. Rev. E 91 (2015) 062115.
\newblock \href {http://dx.doi.org/10.1103/PhysRevE.91.062115}
  {\path{doi:10.1103/PhysRevE.91.062115}}.

\bibitem{flo15}
S.~Florens, I.~Snyman, Universal spatial correlations in the anisotropic kondo
  screening cloud: Analytical insights and numerically exact results from a
  coherent state expansion, Phys. Rev. B 92  195106.
\newblock \href {http://dx.doi.org/10.1103/PhysRevB.92.195106}
  {\path{doi:10.1103/PhysRevB.92.195106}}.

\bibitem{he18}
S.~He, L.~W. Duan, Q.~H. Chen, Improved silbey-harris polaron ansatz for the
  spin-boson model, Phys. Rev. B 97 (2018) 115157.
\newblock \href {http://dx.doi.org/10.1103/PhysRevB.97.115157}
  {\path{doi:10.1103/PhysRevB.97.115157}}.

\bibitem{blu17}
Z.~Blunden-Codd, S.~Bera, B.~Bruognolo, N.~O. Linden, A.~W. Chin, J.~von Delft,
  A.~Nazir, S.~Florens, Anatomy of quantum critical wave functions in
  dissipative impurity problems, Phys. Rev. B 95 (2017) 085104.
\newblock \href {http://dx.doi.org/10.1103/PhysRevB.95.085104}
  {\path{doi:10.1103/PhysRevB.95.085104}}.

\bibitem{zhou15}
N.~J. Zhou, L.~P. Chen, D.~Z. Xu, V.~Chernyak, Y.~Zhao, Symmetry and the
  critical phase of the two-bath spin-boson model: Ground-state properties,
  Phys. Rev. B 91 (2015) 195129.
\newblock \href {http://dx.doi.org/10.1103/PhysRevB.91.195129}
  {\path{doi:10.1103/PhysRevB.91.195129}}.

\bibitem{zhou15b}
N.~J. Zhou, Z.~K. Huang, J.~F. Zhu, V.~Chernyak, Y.~Zhao, Polaron dynamics with
  a multitude of davydov d2 trial states, J. Chem. Phys. 143 (2015) 014113.
\newblock \href {http://dx.doi.org/10.1063/1.4923009}
  {\path{doi:10.1063/1.4923009}}.

\bibitem{zhou16}
N.~J. Zhou, L.~P. Chen, Z.~K. Huang, K.~W. Sun, Y.~Tanimura, Y.~Zhao, Fast,
  accurate simulation of polaron dynamics and multidimensional spectroscopy by
  multiple davydov trial states, J. Phys. Chem. A 120 (2016) 1562--1576.
\newblock \href {http://dx.doi.org/10.1021/acs.jpca.5b12483}
  {\path{doi:10.1021/acs.jpca.5b12483}}.

\bibitem{wan16}
L.~Wang, L.~P. Chen, N.~J. Zhou, Y.~Zhao, Variational dynamics of the sub-ohmic
  spin-boson model on the basis of multiple davydov d1 states, J. Chem. Phys.
  144 (2016) 024101.
\newblock \href {http://dx.doi.org/10.1063/1.4939144}
  {\path{doi:10.1063/1.4939144}}.

\bibitem{wan17}
L.~Wang, Y.~Fujihashi, L.~P. Chen, Y.~Zhao, Finite-temperature time-dependent
  variation with multiple davydov states, J. Chem. Phys. 146 (2017) 124127.
\newblock \href {http://dx.doi.org/10.1063/1.4979017}
  {\path{doi:10.1063/1.4979017}}.

\bibitem{bul03}
R.~Bulla, N.~H. Tong, M.~Vojta, Numerical renormalization group for bosonic
  systems and application to the sub-ohmic spin-boson model, Phys. Rev. Lett.
  91 (2003) 170601.
\newblock \href {http://dx.doi.org/10.1103/PhysRevLett.91.170601}
  {\path{doi:10.1103/PhysRevLett.91.170601}}.

\bibitem{fre13}
M.~F. Frenzel, M.~B. Plenio, Matrix product state representation without
  explicit local hilbert space truncation with applications to the sub-ohmic
  spin-boson model, New J. Phys. 15~(7) (2013) 073046.

\bibitem{kos74}
J.~M. Kosterlitz, The critical properties of the two-dimensional xy model, J.
  Phys. C: Solid State Phys. 7~(6) (1974) 1046.

\bibitem{hur07}
K.~Le~Hur, P.~Doucet-Beaupr\'e, W.~Hofstetter, Entanglement and criticality in
  quantum impurity systems, Phys. Rev. Lett. 99 (2007) 126801.
\newblock \href {http://dx.doi.org/10.1103/PhysRevLett.99.126801}
  {\path{doi:10.1103/PhysRevLett.99.126801}}.

\bibitem{yam19}
T.~Yamamoto, T.~Kato, Microwave scattering in the subohmic spin-boson systems
  of superconducting circuits, J. Phys. Soc. Jpn. 88 (2019) 094601.
\newblock \href {http://dx.doi.org/10.7566/JPSJ.88.094601}
  {\path{doi:10.7566/JPSJ.88.094601}}.

\end{thebibliography}
\bibliographystyle{elsarticle-num}

\end{document}